\documentclass[preprint,12pt]{elsarticle}

\usepackage{amssymb}

\journal{Chemical Physics Letters}

\usepackage{amsmath}

\usepackage{hyperref}

\usepackage{graphicx}

\begin{document}

\begin{frontmatter}



\title{Quantum Revivals of Morse Oscillators and Farey-Ford Geometry}


\author[uark,fhcrc]{Alvason Zhenhua Li}
\ead{alvali@fhcrc.org}

\author[uark]{William G. Harter}
\ead{wharter@uark.edu}

\address[uark]{Microelectronics-Photonics Program, Department of Physics,
University of Arkansas, Fayetteville, AR 72701, USA}
\address[fhcrc]{Present Address: Fred Hutchinson Cancer Research Center, Seattle, WA 98109, USA}

\begin{abstract}
Analytical eigensolutions for Morse oscillators are used to investigate quantum
resonance and revivals and show how Morse anharmonicity affects revival times.
A minimum semi-classical Morse revival time \(T_{min-rev}\) found by Heller is related 
to a complete quantum revival
time \(T_{rev}\) using a quantum deviation \(\delta _N\) parameter that in turn relates \(T_{rev}\) to the maximum quantum beat period \(T_{max-beat}\). Also, number theory of Farey and Thales-circle geometry of Ford is shown to elegantly analyze and display fractional revivals. Such quantum dynamical analysis may have applications for spectroscopy or quantum information processing and computing.
\end{abstract}


\end{frontmatter}

\section{Introduction}
Wavepacket dynamics has a long history that has more recently
been accelerated by graphics that exhibit spacetime behavior. Such studies began with revivals in cavity QED
simulations by Eberly \cite{Eberly:1980} and later simulations of molecular
rovibronic dynamics \cite{Heller:1975, Harter:1982}. Ultrafast
laser spectroscopy made it  possible to observe 
wavepacket resonance and localized periodic motion in experimental
situations \cite{Zewail:1988, Thumm:2003, Ullrich:2006} involving AMOP dynamics \cite{Ullrich:2006, Thumm:2008}. This helped reveal new physics and chemistry of ultrafast spectroscopy \cite{Thumm:2008,Ohmori:2009}.

Some of this involves symmetry and number theoretic properties of wavepacket space-time structure, a still largely
unexplored field. The following development is based upon earlier \(Cn\)-group
and Farey-sum-tree \cite{Farey:1816} analysis of quantum rotors \cite{Harter:2001, Harter:2001Wave}
as cited by Schleich et al. \cite{Schleich:2002, Schleich:2008} for possible numeric factorizing applications. That work
treated only R(2) rings or 1D infinite-wells but nevertheless revealed general symmetry properties.

Here  Morse oscillators are shown to share Farey-sum revival structure of R(2) rings or 1D infinite-wells.
Moreover, Morse revivals
reveal concise ways to find complete revival times \(T_{rev}\) along with new ways to quantify quantum wavepacket dynamics using Ford circles \cite{Ford:1938}\cite{Alvason:2013}.

The Morse oscillator potential \autoref{eq:MorsePotential} is an anharmonic potential \cite{Morse:1929} used to describe covalent molecular bonding. Some dynamics of Morse states have been studied 
\cite{Nieto:1980, Springborg:1988, Levine:1990, Hussin:2008, Heller:2009}
as a model of vibrational anharmonicity.
\begin{subequations}
\begin{equation}
        \label{eq:MorsePotential}
        V_M(x)=D(1-e^{-\alpha x})^2
\end{equation}
Coordinate  \(x\) is  variation of  bond from equilibrium
 where the potential has its minimum and zero value at \(x=0\).
Coefficient
\(D\) is  bond dissociation energy and its maximum inflection value at infinite \(x\).
D relates  harmonic frequency \(\omega_e\) 
 in \autoref{eq:DofMorse} and anharmonic frequency \(\omega_\chi\) in \autoref{eq:aofMorse} that gives width parameter \(\alpha\). The latter is related to reduced mass \(\mu\) and anharmonic frequency
\(\omega_\chi\). 
        \begin{equation}
        \label{eq:DofMorse}
        D=\frac{{\omega_e}^2}{4\omega_\chi}\hbar
        \end{equation}
  
        \begin{equation}
        \label{eq:aofMorse}
        \alpha=\sqrt{\frac{2\omega_{\chi}\mu}{\hbar}}=\sqrt{\frac{\omega{_e}^2\mu}{2D}}
        \end{equation}
\end{subequations}

McCoy \cite{McCoy:2011}  revived interest in exact  eigenfunctions
and eigenvalues \cite{Dong:2002} of Morse oscillator used in \autoref{eq:EnMorse}
and \autoref{eq:WaveMorse} below and allows analysis of its quantum dynamics that may be relevant to anharmonic dynamics in general.

  The Morse oscillator, being anharmonic, has varying spacing
of its energy levels   in contrast to uniform (harmonic) spacing. At high quanta \(n\), energy levels  \(E_n=\hbar \omega_n\)
 have low-\(n\)  spacing \(\Delta E=\hbar
\omega_e\) compressed  for  positive anharmonic frequency
\(\omega_\chi\) in \autoref{eq:EnMorse}.
\begin{equation}
\label{eq:EnMorse}
E_n=\hbar \omega_n=\hbar \omega_e(n+\frac{1}{2})-\hbar \omega_\chi (n+\frac{1}{2})^2
\end{equation}
The corresponding Morse eigenfunctions of the eigenvalues are given by \autoref{eq:WaveMorse}
where \(L_n^{2s}\) represents a generalized associated Laguerre polynomial \cite{McCoy:2011}. 
\begin{subequations}
  \begin{equation}
  \label{eq:WaveMorse}
    \phi_n(x)=e^{\frac{-y(x)}{2}} y(x)^{s(n)} 
    \sqrt{\frac{\alpha(\nu-2n-1)n!}{\Gamma(\nu-n)}}L_n^{2s(n)}(y(x)) 
  \end{equation}
Exponentially scaled \(y(x)\)  has exponent \(s(n)\) as given.
  \begin{equation}
    y(x)=\nu e^{-\alpha x}
  \end{equation}
  \begin{equation}
    s(n)=\frac{1}{2}(\nu-2n-1) 
  \end{equation}
The scaling parameter \(\nu\)  is as follows.  
    \begin{equation}
    \nu=\frac{4D}{\hbar \omega_e}
  \end{equation}
\end{subequations}

Dynamic waves are combinations of eigenfunctions. 
\begin{equation}
  \label{eq:MorseWavePacket}
  \psi(x,t)=\sum_{n=0}^{n_{max}}c_n\phi_n(x)e^{-i\frac{E_nt}{\hbar}} 
\end{equation}
Here \(n_{max}\) is the highest bound state. Its eigenvalue is nearest to dissociative
limit \(D\).  To get maximum beating we assume  equal Fourier coefficients \(c_n=1\). (We do not consider shorter revivals had by zeroing select \(c_{n}\).)  

\begin{figure*}
  \centering
  \includegraphics[width=5.25in]{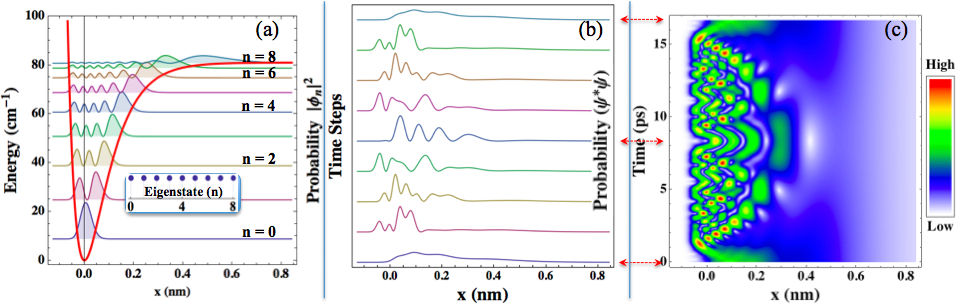}
  \caption{The Morse oscillator with a harmonic frequency \(\omega_e/2\pi c=18(cm^{-1})\)
and an anharmonic frequency \(\omega_\chi/2\pi c=1(cm^{-1})\). (a) Each of its
stationary eigenstate \({|\phi_n}|^2\) was list-plotted on a energy level
of eigenvalue
\(E_n\) in the potential well (red-color-line),
these wave functions are normalized (indicated by the same-height dotted-line).
(b) The wave packet \(\psi^*\psi\)  is propagated along the time steps.
(c) The
probability density map of the wave packet \(|\psi|\) as a function of space
and time.
The double arrows connecting (b)-(c) indicate the corresponding time events.}
\label{fig:WaveMorse}
\end{figure*}

A sample Morse oscillator potential shown in \autoref{fig:WaveMorse}(a) has a total of nine stationary bound states (from \(n=0\) to \(n_{max}=8\)).  The initial wave packet (\autoref{eq:MorseWavePacket} at \(t=0\))
is a sum of these stationary bound states and evolves as shown in \autoref{fig:WaveMorse}(b) ending in its lowest \(\psi(x,T)^{*}\psi(x,T)\) trace as the  initial shape fully revived.

Space-time plots of the norm \(|\psi(x,t)|\) in \autoref{fig:WaveMorse}(c) show  resonant beat nodes and anti-nodes
that outline semi-classical trajectories \(x(t)\)
 corresponding to energy values \(E_n\) ranging from the lowest ground state
\(E_0\) up to the highest bound state \(E_{n_{max}}\).

\section{Analysis}
An essential part of  wave packet dynamics analysis of anharmonic systems
 is to predict if and when exact
wave packet revival might occur. If  \(T_{rev}\) is a time for a Morse
oscillator revival  , then Wang and Heller \cite{Heller:2009} have shown
\begin{subequations}
        \begin{equation}
        \label{eq:GeneralTrevTminRevMorse}
        T_{rev}=\frac{\pi  }{\omega_{\chi}}\mathbb{M}
        \end{equation}
where   \(\mathbb{M}\) is an integer. 
This  revealed two facts about Morse
oscillator dynamics. First, there may be  minimum or fundamental revival period
at 
        \begin{equation}
        \label{eq:TminRevM}
        T_{min-rev}=\frac{\pi}{\omega_{\chi}}
        \end{equation}
\end{subequations}
This is the shortest revival time
for Morse oscillator found by Wang and Heller \cite{Heller:2009}.
Second, any  complete revival period is made of integer
numbers of the fundamental period.
That is,  any complete quantum revival must contain integer numbers
of semiclassical-trajectory-profile periods (minimum revival period) which
is approximately outlined by a classical particle oscillating with a 
frequency of \(2\omega_{\chi}\) in the Morse potential. 

\begin{figure}
  \centering
  \includegraphics[width=5.25in]{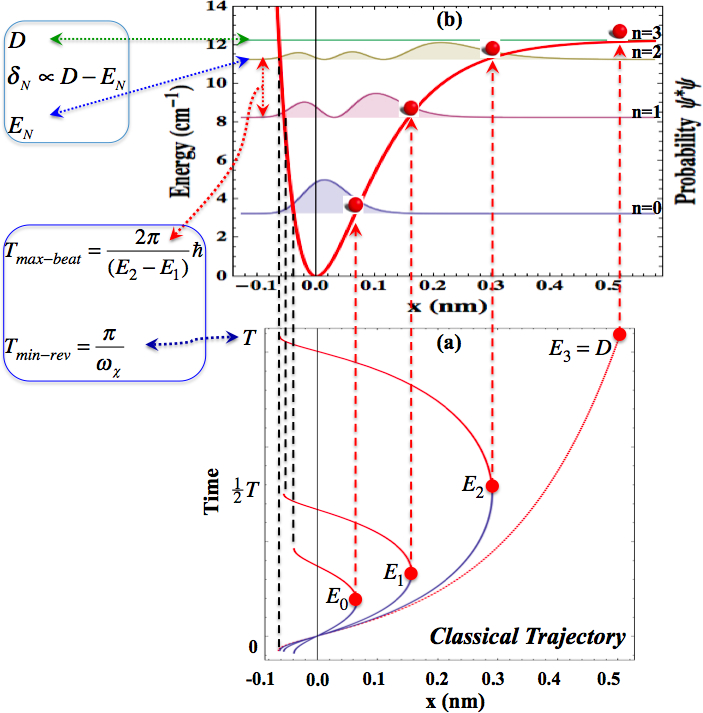}
  \caption{Relating the maximum beat period  and semiclassical-trajectory-profile
period (the minimum revival period). (a) 3 classical trajectories of particles
 oscillating in a Morse potential are plotted in one period time, and one
additional classical trajectory of particle with dissociation energy \(D\)
is also plotted. The red-dots in (a) and (b)  indicate that  these classical
particles have the same energies as the corresponding quantum eigenvalue
energies. (b) The probability amplitudes of 3 bound quantum eigenfunctions
are listed along energy level in a Morse potential (red-thick-line).
  }
  \label{fig:ClassicalOrbital}
\end{figure}

To illustrate relations between quantum periods and  semiclassical-trajectory-profile periods, consider three cases of classical particles with corresponding
quantum eigenvalue energies orbiting in a Morse potential as
shown in \autoref{fig:ClassicalOrbital}(a). Here the rainbow-shape trajectory of
a classical particle with energy  \(E_2\) has a classical oscillating
period \(T\) close to the  fundamental period of \(\pi/\omega_{\chi}\), while
a classical
trajectory with energy \(E_3=D\)  is of a particle barely escaping from
its Morse potential well. 

The preceding case has a simple revival period formula. More analysis
is required to determine a specific integer \(\mathbb{M}\) of  \autoref{eq:GeneralTrevTminRevMorse} for Morse
revivals for a given  \((\omega_e ,\omega_{\chi})\).

Beating of waves with nearby frequency plays a key role in quantum dynamics. The maximum beat period \(T_{max-beat}\) due to the closest
bound energy level pair in the Morse well is one key to finding its revival period.
 A complete revival of \(|\Psi(x,t)|^2\) at time \(T_{rev}\) must contain
integer numbers of all beat periods including at least one
fundamental time period \(T_{max-beat}\) for the slowest beat frequency.
This relates it to revival period.\begin{equation}
\label{eq:GeneralTrevTmaxBeatMorse}
T_{rev}=T_{max-beat}\mathbb{N}
\end{equation}
Here \(\mathbb{N}\) is an integer.
The Morse energy level \autoref{eq:EnMorse} gives a beat-gap between
neighboring energy. 
\begin{equation}
\label{eq:EnergyGapM}
\Delta E= E_{n}-E_{n-1}=\hbar(\omega_{e}-2\omega_{\chi}n)
\end{equation}
The \(\Delta E\) is
the minimum for maximum  \(n\)  occurring  between the highest bound quantum numbers \(n_{max}\) and \(n_{max-1}\).
Planck's relation \(E=\hbar \omega \)
gives maximum
beat period. 
\begin{align}
\label{eq:TmaxBeatM}
T_{max-beat}
=\frac{2\pi}{(\Delta\omega)_{min}}
=\frac{2\pi}{E_{n_{max}}-E_{n_{max-1}}}\hbar
\nonumber\\
=\frac{2\pi}{\omega_{e}-2\omega_{\chi}n_{max}}
\end{align}

 To estimate   upper bound quantum \(n_{max}\) in \autoref{eq:TmaxBeatM},
 we suppose \(n_{max}\) is the integer part of a real number \(n_{real}\) and
 substitute \(n_{real}\) into energy \autoref{eq:EnMorse} to give \(E_{n_{real}}\) that equals  dissociative
limit \(D\) in \autoref{eq:DofMorse}. This equivalent relation is expressed
as
\begin{subequations}
  \begin{align}
  \label{eq:nRealD}
  E_{n_{real}}=\hbar \omega_e(n_{real}+\frac{1}{2})-\hbar\omega_\chi (n_{real}+\frac{1}{2})^2
  \nonumber\\
              =D=\frac{\omega_e^2}{4\omega_\chi}\hbar
  \end{align}
A perfect square equation gives one root.
  \begin{equation}
  \label{eq:nReal}
  n_{real}=\frac{w_e}{2w_\chi}-\frac{1}{2}
  \end{equation}
The integer part or floor of \(n_{real}\) is  the highest
Morse quantum number \(n_{max}\) (For \autoref{fig:WaveMorse}, this is \(n_{max}=8\)).
  \begin{equation}
  \label{eq:nMaxM}
  n_{max}=Floor[n_{real}]=Floor[\frac{w_e}{2w_\chi}-\frac{1}{2}]
  \end{equation}
The fractional part \(\delta_{N}\) of \(n_{real}\) is quantum defect of dissociative
level \(D\)  and  highest allowed bound energy level.
  \begin{equation}
  \label{eq:deltaN}
  \delta_N=n_{real}-n_{max}
  \end{equation}  
\end{subequations}
As illustrated in \autoref{fig:ClassicalOrbital}(b), \(\delta_N\) is proportional
to energy gap between \(D\) and the highest bound energy level. 

Then, the fundamental period \(T_{max-beat}\) in \autoref{eq:TmaxBeatM} is expressed in term of \(\delta_N\) .
\begin{subequations}
        \begin{align}
        \label{eq:TmaxBeatMdN}
  T_{max-beat}
  =\frac{2\pi}{\omega_{e}-2\omega_{\chi}n_{max}}
  =\frac{2\pi}{\omega_{e}-2\omega_{\chi}(n_{real}-\delta_{N})}
  \nonumber\\
  =\frac{2\pi}{\omega_{e}-2\omega_{\chi}(\frac{\omega_e}{2\omega_\chi}-\frac{1}{2}-\delta_{N})}
  \nonumber\\
  =\frac{\pi}{\omega_{\chi}(\delta_N+\frac{1}{2})}
        \end{align}
\autoref{eq:TmaxBeatMdN} is rewritten by substituting \(T_{min-rev}=\pi/\omega_{\chi}\) given by \autoref{eq:TminRevM}.  
        \begin{equation}
        \label{eq:RelationTmaxTmin}
  T_{max-beat}
  =\frac{\pi}{\omega_{\chi}(\delta_N+\frac{1}{2})}
  =T_{min-rev}\frac{1}{(\delta_N+\frac{1}{2})}
        \end{equation}
\end{subequations}
This relates two fundamental building blocks of a
complete Morse revival period.
\begin{subequations}
        \begin{equation}
        \label{eq:RatioTminTmax}
        \frac{T_{min-rev}}{T_{max-beat}}=\delta_{N}+\frac{1}{2}
        \end{equation}
As discussed for \autoref{eq:GeneralTrevTminRevMorse} and
\autoref{eq:GeneralTrevTmaxBeatMorse}, a perfect quantum revival period of
the Morse oscillator \(T_{rev}\) is  composed of  integer numbers of the
fundamental periods as follows. 
        \begin{equation}
        \label{eq:TrevTminTmax}
        T_{rev}=T_{min-rev}\mathbb{M}=T_{max-beat}\mathbb{N}
        \end{equation}
Then \autoref{eq:RatioTminTmax} and \autoref{eq:TrevTminTmax} relate  N and M integers.
        \begin{equation}
        \frac{\mathbb{N}}{\mathbb{M}}=\delta_{N}+\frac{1}{2}
        \end{equation}
\end{subequations}
  
The quantum  beat-period approach gives Morse  revival
time \(T_{rev}\)  in terms of \(T_{max-beat}\)
and \(\delta_N\) as follows.
\begin{subequations}
        \begin{equation}
        \label{eq:TrevTmax}
        T_{rev}=T_{max-beat}\mathbb{N}=T_{max-beat}Numerator[\delta_N + \frac{1}{2}]
        \end{equation}
The semiclassical-trajectory-profile approach gives
\(T_{rev}\)  in terms of \(T_{min-rev}\) and \(\delta_N\)
as follows.
        \begin{equation}
        \label{eq:TrevTmin}
        T_{rev}=T_{min-rev}\mathbb{M}=T_{min-rev}Denominator[\delta_N + \frac{1}{2}]
        \end{equation}
\end{subequations}

Both \(T_{min-rev}\) or \(T_{max-beat}\) serve as a fundamental
building blocks of \(T_{rev}\). Examples of this follow.

\subsection{Fibonacci Sequence and Exchange Rate of \(\frac{T_{min-rev}}{T_{max-beat}}\)}
\begin{figure*}
  \centering
  \includegraphics[width=5.25in]{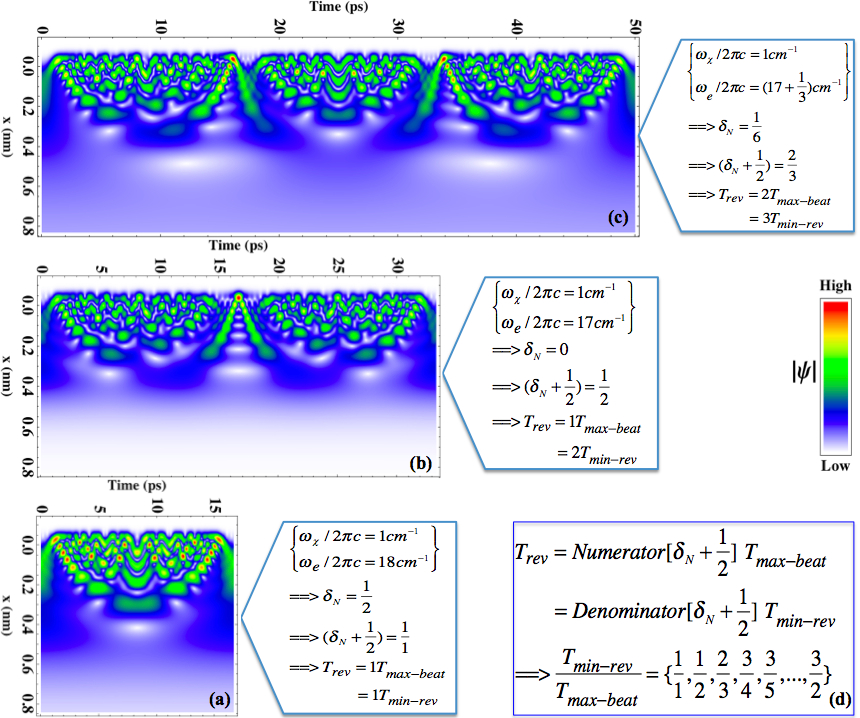}
  \caption{Fibonacci sequence and exchange rate.
(a) When \(\delta_N=1/2\),
the \(T_{rev}\) is composed of one \(T_{max-beat}\) and one \(T_{min-rev}\).
(b) When \(\delta_N=0\),
the \(T_{rev}\) is composed of \(1T_{max-beat}\) and \(2T_{min-rev}\). (c)
When \(\delta_N=1/6\),
the \(T_{rev}\) is composed of  \(2T_{max-beat}\) and \(3T_{min-rev}\). (d)
Allowed ratios of \(T_{min-rev}\)
to \(T_{max-beat}\). 
}
  \label{fig:FibonacciMorse}
\end{figure*}

Interplay of harmonicity and anharmonicity of Morse
oscillators affects revival period \(T_{rev}\). Consider \autoref{fig:FibonacciMorse}(a)-(c) where the value of \(T_{rev}\)
is increased  from  the minimum revival
period \(T_{min-rev}\) to multiples thereof with fixed  anharmonic frequency \(w_\chi/2\pi c = 1(cm^{-1})\). 

In \autoref{fig:FibonacciMorse}(a) with \(\omega_e/2\pi
c =18(cm^{-1})\)  one perfect revival occurs in the minimum revival time:  \(T_{rev} = T_{max-beat} = T_{min-rev}\)
giving a unit ratio  \(T_{min-rev}/T_{max-beat} = 1/1\).

Then in \autoref{fig:FibonacciMorse}(b) with \(\omega_e/2\pi c = 17(cm^{-1})\) and the same \(\omega_{\chi}\), is seen a double time for perfect revival  of
\(T_{rev} = T_{max-beat} = 2T_{min-rev}\) with  a half ratio  \(T_{min-rev}/T_{max-beat} = 1/2\).
We note that this double revival time \(T_{rev} = 2T_{min-rev} = 2\pi/\omega_\chi\)
exactly equals  \(T_{approx}\) in \autoref{eq:TapproxM} given by a semiclassical treatment of general anharmonic oscillators \cite{Perelman:1989,
Robinett:2004, Engel:2004, Umanskii:1994} that
assumes large quantum numbers \(n\) around their average \(\bar{n}\).
\begin{equation}
  \label{eq:TapproxM}
  T_{approx}=\frac{2\pi}{\frac{1}{2}\left\vert\frac{d^2E_n}{dn^2}\right\vert_{n=\bar
  n}} = \frac{2\pi}{w_\chi}
\end{equation}

In \autoref{fig:FibonacciMorse}(c) with \(w_e/2\pi c = 17+\frac{1}{3}(cm^{-1})
  \) is a perfect revival time  \(T_{rev} = 2T_{max-beat} = 3T_{min-rev}\) with ratio of minimum revival period to  maximum beat period
of \(T_{min-rev}/T_{max-beat} = 2/3\).
 
All revival periods \(T_{rev}\) are composed of an
integer number of fundamental period \(T_{min-rev}\) (or \(T_{max-beat}\)) but have a differing ratios \(T_{min-rev}/T_{max-beat}\) that range between \(1/2\) and \(3/2\). The Fibonacci
sequence \(\{1/1, 1/2, 2/3, 3/5, 5/8, ....\}\) is a subset of the possible rational ratios \(T_{min-rev}/T_{max-beat}\). 

\subsection{Farey-sum and Ford geometry of fractional revivals}

\begin{figure*}
  \centering
  \includegraphics[width=5.25in]{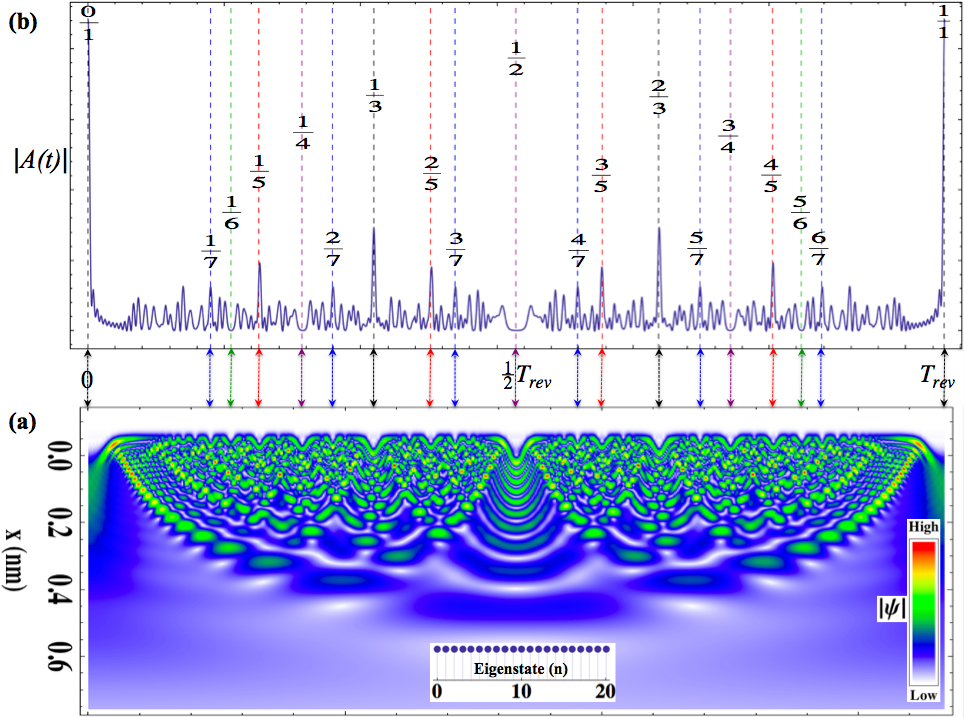}
  \caption{The Farey-sum sequence structure appears in Morse oscillator space-time
pattern for \(w_e/2\pi c = 42(cm^{-1})\) and
\(w_\chi/2\pi c = 1(cm^{-1})\) with \(n_{max}=20\). (a) One complete revival period plot
of the wave packet has color
denote magnitude \(\left\vert \Psi(x,t) \right\vert\).
(b) The norm of autocorrelation
 function
  \((\left\vert A(t) \right\vert\) with \(n_{max}=20)\) is plotted in one
complete revival period \(T_{rev}\) whose fractions
\(\{
\frac{0}{1},
\frac{1}{7},\frac{1}{6},\frac{1}{5},\frac{1}{4},
\frac{2}{7},\frac{1}{3},\frac{2}{5},
\frac{3}{7},\frac{1}{2},
\frac{4}{7},\frac{3}{5},\frac{2}{3},
\frac{5}{7},\frac{3}{4},\frac{4}{5},\frac{5}{6},
\frac{6}{7},
\frac{1}{1}
\}\) 
are denoted by the vertical dashed lines.  The double arrows
connecting (a)-(b) indicate the corresponding time events having peaks (or nodes) for time fraction \(\frac{n}{d}\) of odd (or even) depth \(d\).}
  \label{fig:FareyMorse}
\end{figure*}
 
Fractional or intra-revival structure of  Morse vibrators is quite like that of  rotor revivals. In \autoref{fig:FareyMorse}(a) is a  Morse revival of higher frequency
\(\omega_e/2\pi c = 42(cm^{-1})\) and more states \((n_{max}
= 20)\)  than the one in \autoref{fig:FibonacciMorse}(a), but with the same anharmonicity \(\omega_\chi/2\pi c = 1(cm^{-1})\) and revival period: 
\(T_{rev}=1T_{min-rev}=1T_{max-beat}=1/(2c(cm)^{-1})\approx16.7(picro-second)\).

Fractional revival structure is  visible as a series of dips on top of \autoref{fig:FareyMorse}(a) and in Fourier amplitude frequency  \(\omega_n\)  sum or autocorrelation \(A(t)\)   spectra \cite{Robinett:2004} in \autoref{fig:FareyMorse}(b).
\begin{equation}
\label{eq:AutoCorrelation}
  A(t) = \sum_{n=0}^{n_{max}}e^{-i\frac{E_nt}{\hbar}} 
       = \sum_{n=0}^{n_{max}}e^{-i\omega_nt}
\end{equation}

\(|A(t)|\) spectra match Farey-sum sequence used in 1815 by geologist John Farey \cite{Farey:1816}\cite{Hardy:1979} to analyze tidal beats. 

A Farey sequence, starting with fraction
\(0/1\) and ending with fraction \(1/1\), builds  hierarchies of irreducible rational fractions on a real line between \(0.0\) and \(1.0\) \cite{Hardy:1979}. 
A Farey-sum  \(\frac{a}{b}\)+\(\frac{c}{d}\)=\(\frac{a+c}{b+d}\) is a curious process to locate significant fractions \(\frac{n}{d}\) between \(\frac{a}{b}\) and \(\frac{c}{d}\)
or overtone \((n:d)\) resonances in between an \((a:b)\) and a \((c:d)\) resonance. 

\begin{figure}
\centering
  \includegraphics[width=5.25in]{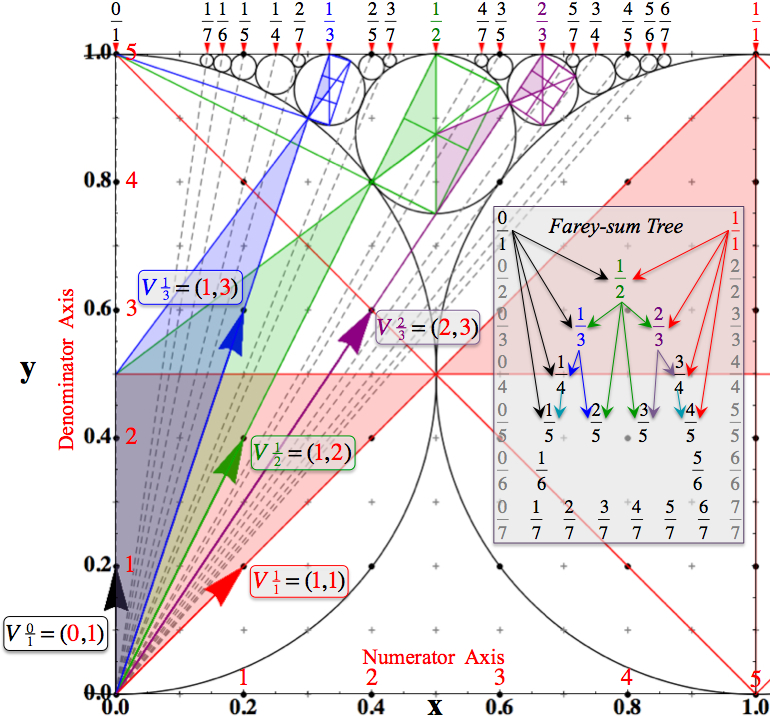}
  \caption{Ford circles and vectors with Farey-sum
sequence.}
  \label{fig:FordCircles}
\end{figure}

In 1938, Leslie Ford \cite{Ford:1938} found a geometric description that helps elucidate Farey-sums. Ford geometry views each fraction as a vector and the Farey sum as a vector sum in Denominator\((y)\)-vs-Numerator\((x)\) space such as is plotted for \(0\leq(x,y)\leq1\) in \autoref{fig:FordCircles}. A fraction  \(\frac{a}{b}\) is drawn as a vector with tail at origin and head at the point \((x=a,y=b)\) as shown by examples \(\mathbf{V}_\frac{0}{1}\) and \(\mathbf{V_\frac{1}{1}}\), the black and red arrows in \autoref{fig:FordCircles} (lower left).
\begin{subequations}
\begin{equation}
    \mathbf{V_\frac{0}{1} + V_\frac{1}{1}}
     =\begin{pmatrix}0 \\1 \\ \end{pmatrix}+
      \begin{pmatrix}1 \\1 \\ \end{pmatrix}
     =\begin{pmatrix}1 \\2 \\ \end{pmatrix}
     =\mathbf{V_\frac{1}{2}}
    \label{eq:FordVectorSum1}
  \end{equation}
The green arrow in \autoref{fig:FordCircles} is \(\mathbf{V_\frac{1}{2}}\). Each vector \(\mathbf{V_{\frac{n}{d}}}\) points to or intersects a real value \(x=n/d \) on the top \((y=1)\)-line of \autoref{fig:FordCircles}. Ford \cite{Ford:1938} discovered that each 
\(x = n/d\) is a   tangent point  for a circle having diameter \(1/d^{2}\) hanging  below the top \((y=1)\)-line that is itself tangent to  infinite sequences of smaller such circles, each  tangent to the next and converging on \(x = n/d\). The \((d=1)\)-Ford-circle is a unit-diameter circle sliced to fit the unit \((x,y)\)-area with a pair of tangent semi circles belonging to unit Ford base vectors \(\mathbf{V}_\frac{0}{1}\) and  \(\mathbf{V}_\frac{1}{1}\). Fractions \(\frac{0}{1}\) and \(\frac{1}{1}\) make a second (\(d\)=2)-Ford\(\) circle of diameter \(1/2^2\) in the upper center of \autoref{fig:FordCircles} pointed out by sum 
\(\mathbf{V}_\frac{1}{2}\)=\(\mathbf{V_\frac{0}{1}}\)+\(\mathbf{V_\frac{1}{1}}\)
in \autoref{eq:FordVectorSum1}. 

It is tangent to  ``parent" Ford circles for \(\frac{0}{1}\) and \(\frac{1}{1}\).
Also shown is (\(d\)=3)-Ford\(\) circle for vector
\(\mathbf{V_\frac{1}{3}}\)=\(\mathbf{V_\frac{0}{1}}\)+\(\mathbf{V_\frac{1}{2}}\) that is tangent to circles of
its parent fractions \(\frac{0}{1}\) and \(\frac{1}{2}\). 
\begin{equation}
    \mathbf{V_\frac{0}{1} + V_\frac{1}{2}}
     =\begin{pmatrix}0 \\1 \\ \end{pmatrix}+
      \begin{pmatrix}1 \\2 \\ \end{pmatrix}
     =\begin{pmatrix}1 \\3 \\ \end{pmatrix}
     =\mathbf{V_\frac{1}{3}}
    \label{eq:FordVectorSum2}
  \end{equation}
\end{subequations}

\begin{figure}
\centering
  \includegraphics[width=5.25in]{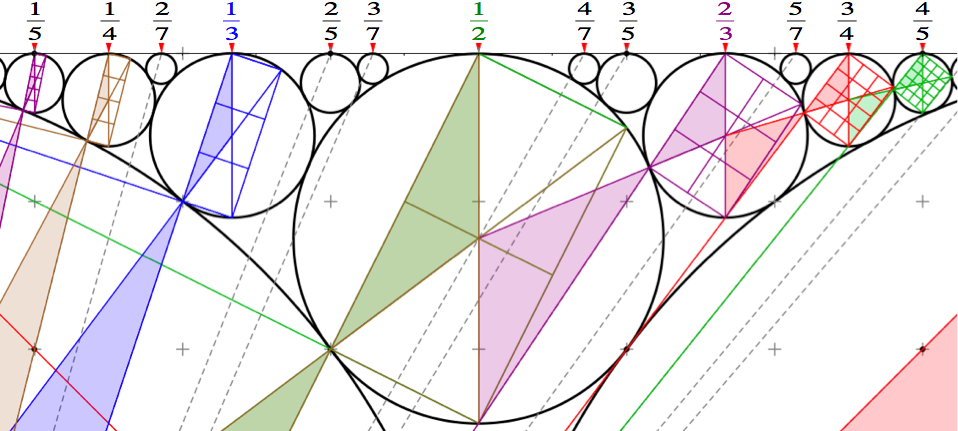}
  \caption{``Quantization'' inside Ford circles: 
pixel lattices of 
\{\textit{(1\textit{\emph{x}}5)}, \textit{(1\textit{\emph{x}}4)}, \textit{(1\textit{\emph{x}}3)},
\textit{(1\textit{\emph{x}}2)}, \textit{(2\textit{\emph{x}}3)}, \textit{(3\textit{\emph{x}}4)}, \textit{(4\textit{\emph{x}}5)}\}
rectangles lie circumscribed by circles of fractions
\(\{
\frac{1}{5},\frac{1}{4},
\frac{1}{3},\frac{1}{2},\frac{2}{3},
\frac{3}{4},\frac{4}{5}
\}\).}
  \label{fig:InsideFordCircles}
\end{figure}  

Thales \textit{600-BCE} rectangle-in-circle geometry is sufficient to derive Ford geometry. Tangent Ford circles like the \(\frac{0}{1}\) , \(\frac{1}{1}\), and \(\frac{1}{2}\) circles in \autoref{fig:FordCircles} meet at corners of similar Thales rectangles whose vertical diagonals are circle diameters hanging below their respective fraction points.  Circle diameters
subtend \(90^{\circ}\) corners by Thales theorem. The \(\frac{0}{1}\) and \(\frac{1}{2}\) corners meet where the \(\mathbf{V_\frac{1}{2}}\)-vector line crosses the \(\frac{0}{1}\)-circle. This is the \(\frac{0}{1}\)-\(\frac{1}{2}\)-circle-tangent point.  A \(\frac{0}{1}\)-diameter line through that point intersects the vertical diameter of the \(\frac{1}{2}\)-circle at its center thus defining it. Similar geometry (not drawn) applies to  the \(\frac{1}{2}\)-\(\frac{1}{1}\)-circle-tangent. 
A Farey-sum of a circle-tangent pair is a  new Ford circle and fraction as shown by examples \(\frac{0}{1}\)+\(\frac{1}{2}\)=\(\frac{1}{3}\)  in \autoref{eq:FordVectorSum2} or \(\frac{1}{2}\)+\(\frac{1}{1}\)=\(\frac{2}{3}\) listed in level-3 of  the Farey-sum-tree on the righthand side of \autoref{fig:FordCircles}. Farey sums that give reducible fractions  \(\frac{N}{D}\)=\(\frac{n\cdot f}{d\cdot f}\) are labeled by  their reduced form \(\frac{n}{d}\) with the shortest allowed Ford vector, least  depth or denominator \(d\), and largest possible Ford circle.      

Continued Farey-sums of Ford vectors
give sequences of  circles each belonging to an irreducible fraction \(\frac{n}{d}\) and its vector \(\mathbf{V}_\frac{n}{d}\). In \autoref{fig:FordCircles}, these circles nest in the area between their
original Farey
``grandparent'' circles \(\mathbf{V_\frac{0}{1}}\) and \(\mathbf{V_\frac{1}{1}}\).

A Farey-sum-tree of  fractions of depth
 \(d\leqslant{7}\)
\(\{
\frac{0}{1},
\frac{1}{7},\frac{1}{6},\frac{1}{5},\frac{1}{4},
\frac{2}{7},\frac{1}{3},\frac{2}{5},
\frac{3}{7},\frac{1}{2},
\frac{4}{7},\frac{3}{5},\frac{2}{3},
\frac{5}{7},\frac{3}{4},\frac{4}{5},\frac{5}{6},
\frac{6}{7},
\frac{1}{1}
\}\) 
is shown in box on the right of \autoref{fig:FordCircles} and represented by
a total of \(19\) sequentially and mutually tangent circles.

A revealing portrait emerges of quantum ``fractal'' structure filling the area below the top line with ever tinier \(\frac{1}{d^2}\)-diamter circles as spectral depth \(d\) increases. 

By construction all Ford-vector and Thales-rectangle slopes are rational, but surprisingly so are their aspect ratios ``quantized'' into \(n\)-by-\(d\) pixel arrays. For example,   
\textit{(1\textit{\emph{x}}3)},\textit{(1\textit{\emph{x}}2)}, and\textit{(2\textit{\emph{x}}3)}  pixel arrays lie  inside 
\(\{\frac{1}{3},\frac{1}{2},\frac{2}{3}\}\) circles in \autoref{fig:FordCircles}, and \autoref{fig:InsideFordCircles} shows finer (\textit{n}\textit{\emph{x}}\textit{d}) arrays of pixel rectangles circumscribed by \(\frac{n}{d}\)-circles. 

\section{Conclusion}
In conclusion, exact Morse oscillator
eigensolutions allow  more detailed analysis of their quantum dynamics. A key top-level-to-dissociation gap parameter \(\delta_N\) provides
a concise   revival time formula in terms
of two fundamental periods, a semiclassical \(T_{min-rev}\) found by Wang and Heller and a longest quantum beat period  \(T_{max-beat}\). This     shows that complete revival periods may be composed  of integer numbers of the two. Finally,  fractional revivals seen in rotor pulse evolution is also shown to be present  in Morse wave dynamics in the form of Farey-sum spectral sub-structure. A Ford-circles geometry  relating rational fractions to real numbers may be developed to visualize these phenomena and may eventually have application to
quantum information processing and computing.





\bibliographystyle{elsarticle-num} 


\bibliography{ReferenceMorse}

\begin{thebibliography}{10}
\expandafter\ifx\csname url\endcsname\relax
  \def\url#1{\texttt{#1}}\fi
\expandafter\ifx\csname urlprefix\endcsname\relax\def\urlprefix{URL }\fi
\expandafter\ifx\csname href\endcsname\relax
  \def\href#1#2{#2} \def\path#1{#1}\fi

\bibitem{Eberly:1980}
J.~H. Eberly, N.~B. Narozhny, J.~J. Sanchez-Mondragon, Periodic spontaneous
  collapse and revival in a simple quantum model, Phys. Rev. Lett. 44 (1980)
  1323.

\bibitem{Heller:1975}
E.~J. Heller, Time dependent approach to semiclassical dynamics, J. Chem. Phys.
  62 (1975) 1544.

\bibitem{Harter:1982}
R.~S. McDowell, C.~W. Patterson, W.~G. Harter, The modern revolution in
  infrared spectroscopy, Los Alamos Science 3 (1982) 38.

\bibitem{Zewail:1988}
A.~H. Zewail, Laser femtochemistry, Science 242 (1988) 1645.

\bibitem{Thumm:2003}
B.~Feuerstein, U.~Thumm, Mapping of coherent and decohering nuclear wave-packet
  dynamics in d2 with ultrashort laser pulses, Phys. Rev. A 67~(063408).

\bibitem{Ullrich:2006}
A.~Rudenko, T.~Ergler, B.~Feuerstein, K.~Zrost, C.~D. Schroter, R.~Moshammer,
  J.~Ullrich, Real-time observation of vibrational revival in the fastest
  molecular system, Chem. Phys. 329 (2006) 193--202.

\bibitem{Thumm:2008}
T.~Niederhausen, U.~Thumm, Controlled vibrational quenching of nuclear wave
  packets in d2, Phys. Rev. A 77~(013407).

\bibitem{Ohmori:2009}
K.~Ohmori, Wave-packet and coherent control dynamics, Annu. Rev. Phys. Chem. 60
  (2009) 487--511.

\bibitem{Farey:1816}
J.~Farey, Philos. Mag. 47 (1816) 385.

\bibitem{Harter:2001}
W.~G. Harter, Quantum-fractal revival structure in cn quadratic spectra: Base-n
  quantum computer registers, Phys. Rev. A 64~(012312).

\bibitem{Harter:2001Wave}
W.~G. Harter, Wave node dynamics and revival symmetry in quantum rotors, J.
  Mol. Spectrosc. 210 (2001) 166.

\bibitem{Schleich:2002}
H.~Mack, M.~Bienert, F.~Haug, M.~Freyberger, W.~Schleich, Wave packets can
  factorize numbers, phys. stat. sol. (b) 233~(3) (2002) 408--415.

\bibitem{Schleich:2008}
M.~Gilowski, T.~Wendrich, T.~M{\"u}ller, C.~Jentsch, W.~Ertmer, E.~M. Rasel,
  W.~P. Schleich, Gauss sum factorization with cold atoms, Phys. Rev. Lett.
  100~(030201).

\bibitem{Ford:1938}
L.~R. Ford, Fractions, The American Mathematical Monthly 45~(9) (1938)
  586--601.

\bibitem{Alvason:2013}
A.~Z. Li, Quantum resonant beats and revivals in the morse oscillators and
  rotors, Ph.D. thesis, University of Arkansas (2013).

\bibitem{Morse:1929}
P.~M. Morse, Diatomic molecules according to the wave mechanics. ii.
  vibrational levels, Phys. Rev. 34 (1929) 57.

\bibitem{Nieto:1980}
V.~P. Gutschick, M.~M. Nieto, Coherent states for general potentials. v. time
  evolution, Phys. Rev. D 22 (1980) 403.

\bibitem{Springborg:1988}
J.~P. Dahl, M.~Springborg, The morse oscillator in position space, momentum
  space, and phase space, J. Chem. Phys. 88 (1988) 4535.

\bibitem{Levine:1990}
S.~Kais, R.~D. Levine, Coherent states for the morse oscillator, Phys. Rev. A
  41 (1990) 2301.

\bibitem{Hussin:2008}
M.~Angelova, V.~Hussin, Generalized and gaussian coherent states for the morse
  potential, J. Phys. A: Math. Theor. 41~(304016).

\bibitem{Heller:2009}
Z.~Wang, E.~J. Heller, Semiclassical investigation of the revival phenomena in
  a one-dimensional system, J. Phys. A: Math. Theor. 42~(285304).

\bibitem{McCoy:2011}
A.~B. McCoy, Curious properties of the morse oscillator, Chem. Phys. Lett. 501
  (2011) 603--607.

\bibitem{Dong:2002}
S.~Dong, R.~Lemus, A.~Frank, Ladder operators for the morse potential, Int. J.
  Quantum Chem. 86 (2002) 433.

\bibitem{Perelman:1989}
I.~S. Averbukh, N.~F. Perelman, Fractional revivals universality in the
  long-term evolution of quantum wave packets beyond the correspondence
  principle dynamics, Phys. Lett. A 139 (1989) 449.

\bibitem{Robinett:2004}
R.~W. Robinett, Quantum wave packet revivals, Phys. Rep. 392 (2004) 1--119.

\bibitem{Engel:2004}
T.~Lohmuller, V.~Engel, J.~Beswick, C.~Meier, Fractional revivals in the
  rovibrational motion of i2, J. Chem. Phys. 120~(22) (2004) 10442.

\bibitem{Umanskii:1994}
V.~V. Eryomin, S.~I. Vetchinkin, I.~M. Umanskii, Manifestations of wave packet
  fractional revivals in a morselike anharmonic system, J. Chem. Phys. 101
  (1994) 10730.

\bibitem{Hardy:1979}
G.~H. Hardy, E.~M. Wright, An Introduction to the Theory of Numbers, 5th
  Edition, Oxford University Press, New York, 1979.

\end{thebibliography}






\end{document}